\documentclass{article}
\usepackage{caption}
\usepackage[margin=0.5in]{geometry}
\usepackage{fancyhdr}
\relax  
 
 \renewenvironment{author} {
 \large
 \emph
  }

\newcommand{\keywords}[1]{
\medskip
Keywords: \textit{#1}
}

\newcommand{\dedication}[1]{
\medskip
\textit{#1}
}

\renewenvironment{abstract}{
\small
\medskip\medskip
}

\newcommand{\upcite}[1]{$^{\mbox{\scriptsize \cite{#1}}}$}
\usepackage{graphicx}
\usepackage[singlelinecheck=off,justification=raggedright]{subcaption}
\usepackage{float}
\usepackage{indentfirst}
\usepackage[graphicx]{realboxes}
\usepackage[colorlinks,linkcolor=blue,anchorcolor=blue,citecolor=blue,urlcolor=blue]{hyperref}
\usepackage{amsthm,amssymb}
\usepackage{amsmath}
\usepackage{mathrsfs}
\newdimen\Lmargin
\newdimen\Rmargin
\Rmargin=\paperwidth \advance\Rmargin by-\Lmargin \advance\Rmargin by-\textwidth
\ifdim\Lmargin>\Rmargin \Rmargin=\Lmargin \fi
\def\Scentering{\leftskip=0pt plus 1fil minus \Rmargin
                \rightskip=\leftskip}
\usepackage{cite}
\usepackage{orcidlink}
\newcommand{\orcid}[1]{\href{https://orcid.org/#1}{\includegraphics[width=10pt]{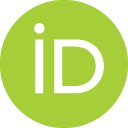}}}
\usepackage{ragged2e}
\usepackage{lineno}

\usepackage{ulem}

\begin{document}
\date{}

\title{Microwave-coupled optical bistability in driven and interacting Rydberg gases}
\maketitle

\vspace{-1cm}

\justifying
\noindent\author{Zhehua Zhang$^{1}$,\ Zeyan Zhang$^{1}$, \ Shaoxing Han$^{1}$,\  Yuqing Zhang$^{1}$,\ Guoqing Zhang$^{1}$ ,\ Jizhou Wu$^{1,2,3\dagger}$~\orcid{0000-0003-0299-8938},\ Vladimir B. Sovkov$^{4}$,\ Wenliang Liu$^{1,2}$,\  Yuqing Li$^{1,2,3}$,\ Linjie Zhang$^{1,2,3}$~\orcid{0000-0001-8668-1895},\ Liantuan Xiao$^{1,2}$,\ Suotang Jia$^{1,2}$,\ Weibin Li$^{5}$~\orcid{0000-0001-6731-1311},\ Jie Ma$^{1,2,3\dagger}$~\orcid{0000-0003-4932-9205}}

~\\
\noindent$^{1}$ State Key Laboratory of Quantum Optics and Quantum Optics Devices, Institute of Laser Spectroscopy, College of Physics and Electronics Engineering, Shanxi University, Taiyuan 030006, China 

\noindent$^{2}$ Collaborative Innovation Center of Extreme Optics, Shanxi University, Taiyuan 030006, China 

\noindent$^{3}$ Hefei National Laboratory, Hefei 230088, China

\noindent$^{4}$ St. Petersburg State University, 7/9 Universitetskaya nab., St. Petersburg 199034, Russia

\noindent$^{5}$ School of Physics and Astronomy and Centre for the Mathematics and Theoretical Physics of Quantum Non-Equilibrium Systems, University of Nottingham, Nottingham NG7 2RD, United Kingdom

\noindent$^{\dagger}$ wujz@sxu.edu.cn; mj@sxu.edu.cn

\dedication{}



\noindent\keywords{Optical bistability, Rydberg atom, Nonequilibrium phenomena, Microwave measurement}

\begin{abstract}
\justifying  
\noindent
Nonequilibrium dynamics are closely related to various fields of research, in which vastly different phases emerge when parameters are changed. However, it is difficult to construct nonequilibrium systems that have sufficiently tunable controllable parameters. Using microwave field coupling induced optical bistability, Rydberg gases exhibit a range of significantly different optical responses. In conjunction with electromagnetically induced transparency, the microwave coupling can create versatile nonequilibrium dynamics. In particular, the microwave coupling of two Rydberg states provides an additional handle for controlling the dynamics. And the microwave-controlled nonequilibrium phase transition has the potential to be applied in microwave field measurement. This study opens a new avenue to exploring bistable dynamics using microwave-coupled Rydberg gases, and developing quantum technological applications.
\end{abstract}


\vspace{1cm}

\justifying 
\noindent
Nonequilibrium processes in complex dynamical systems give rise to a wide range of methodologically instructive and practically promising phenomena, not only in the realm of physics but also across disciplines such as geological sciences, biology, and socioeconomics \upcite{1}. In particular, it holds a pivotal position in contemporary physical researches such as quantum many-body physics, atomic and molecular physics, quantum information, and quantum computing \upcite{2,3,4}. In the pursuit of comprehending nonequilibrium phenomena, it is essential to establish experimental setups that induce nonequilibrium dynamics and phase transitions \upcite{5,6,7,8}.

The phenomenon of optical bistability \upcite{9,10,11,12,13,14}, induced by optical excitation, is a typical nonequilibrium effect arising from nonlinear and competing processes. In atomic systems, the interactions between atoms are generally too weak to significantly impact the nonlinear dynamics of excitation. Previous experiments often relied on amplifying optical fields to enhance nonlinear effects \upcite{15}. Various systems, such as nonlinear prisms \upcite{16}, Quantum Electrodynamics(QED) cavities \upcite{17}, and photonic crystal cavities \upcite{18}, have been utilized to investigate the bistability. The intrinsic optical bistability \upcite{19} formed by dipole-dipole interactions \upcite{20} is particularly challenging to achieve. Additionally, the broadening of the involved atomic levels inhibits the emergence of bistability \upcite{21,22}.

Rydberg atoms serve as a good candidate to investigate the optical bistability \upcite{23,24,25,26,27}. These atoms possess exceptionally strong atomic interactions that can be adjusted by selecting states with different principal quantum numbers or orbital angular moments. In alkali Rydberg atoms, the frequency difference between Rydberg states ranges from 1 MHz to 1 THz, enabling these atoms to function as electric field sensors  \upcite{28,29}. Additionally, Rydberg atoms exhibit strong inter-particle interactions comparable to the dissipation required by the system \upcite{30}. This leads to diverse nonequilibrium dynamics and the formation of numerous dynamic phases \upcite{31,32,33,34}. The intrinsic optical bistability in Rydberg atomic gases was first demonstrated at room temperature in 2013 \upcite{5}. In a subsequent study, D. Ding $et$ $al$. observed optical bistability using electromagnetically induced transparency (EIT) \upcite{35,36} in a heated Rb cell in 2020, where they also investigated self-organized dynamics near the phase transition critical point and developed a model for this behavior based on Rydberg systems \upcite{15}. However, previous studies have primarily focused on observing bistability and studying nonequilibrium phase transitions using a single Rydberg state structure \upcite{5,15,37}. It is important to note that while these simplified models provide valuable insights, they may not fully capture the complexity of nonequilibrium phenomena observed in more sophisticated scenarios.

Unlike previous researches aforementioned, in addition to using EIT for non-destructive detection of bistability, we introduced microwaves as external perturbations. By controlling the microwave frequency and intensity, we can exert a controllable influence on the nonequilibrium phenomena of the system. This suggests that microwaves can serve as stable external disturbances to simulate a broader range of nonequilibrium phenomena. We have illustrated an epidemic model in the supplementary information to support this assertion. Moreover, the information carried by microwave fields is also embedded in the nonequilibrium phenomena, which opens up the possibility of further quantum measurements and quantum communication in nonequilibrium systems.

\section*{RESULTS}
\subsection*{Bistability in EIT}

\begin{figure}
    \Scentering
    \includegraphics[width=1.1 \linewidth]{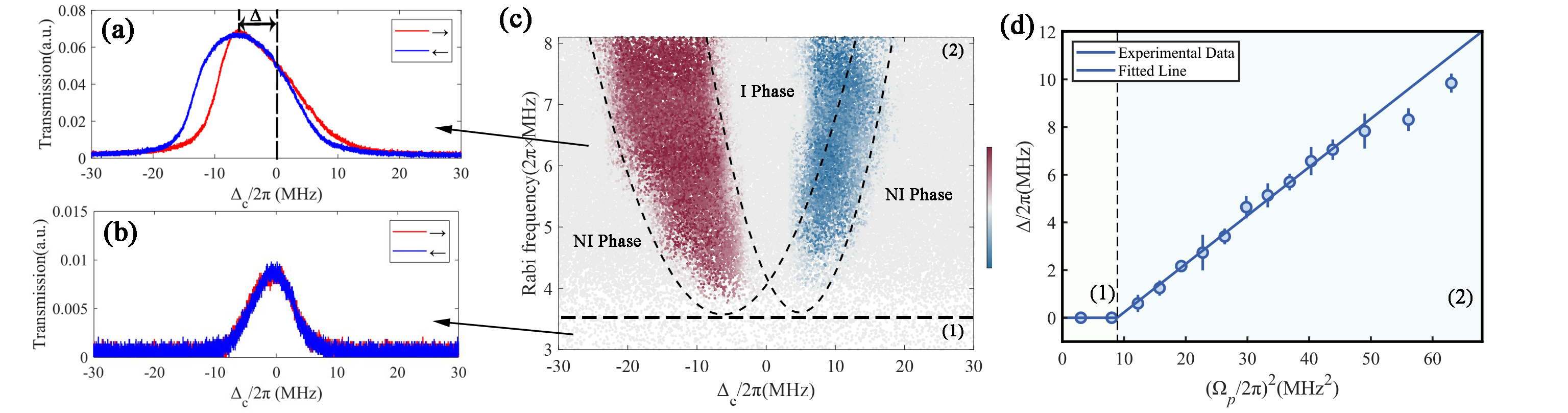} 
    \caption{(a) Transmission spectrum of the probe beam while the coupling beam($\Delta_c$) is scanned from red to blue detuning (red) and vice versa (blue). The probe beam Rabi frequency set to 6.48×2$\pi$ MHz. The bistable phenomenon is clearly seen in the graph. (b) EIT lines without bistable phenomena, observed with the Rabi frequency of 3.24×2$\pi$ MHz. (c) Color map of the difference in probe light transmission when $\Delta_c$ is scanned in different directions (red if transmission is higher for scans from negative to positive $\Delta_c$, blue in the opposite case). (d) Relationship between the shift of peak transmission $\Delta$ and the Rabi frequency of the probe beam. The error bars represent the standard deviation of the data. The two regions in the figure correspond to the two regions in (c). }
    \label{Fig.1}
\end{figure}

\noindent We initially investigate bistability phenomenon in a ladder-type EIT system. The bistable information is acquired through the observation of the transmission spectrum. The experimental setup is shown in Methods section. By scanning forward and backward around the resonance frequency of the coupling beam, the EIT spectra in both directions are clearly visualized, and they form a bistable pattern as shown in Fig.\ref{Fig.1}(a). The EIT spectra generated by forward scanning are shown in red, while those produced by backward scanning are shown in blue. This clear separation in the parameter space indicates the formation of a bistability. The phenomenon of nonequilibrium occurs when the Rabi frequency of the probe laser field is high enough to excite a density of Rydberg atoms in the vapor cell that exceeds the critical density \upcite{5}. We observe that the two spectral lines are non-coincident at either the red or blue detuning, and this effect is more pronounced at the red detuning side. We note that a peak in the forward scan is higher than in the backward scanning spectra, and it gradually disappears as the Rabi frequency of the probe laser field decreases. Although the spectra in both directions deviate from the typical EIT spectra, the spectra from the backward scan exhibit more Gaussian characteristics and do not have a sharp peak. Therefore, we use the red-detuning region of the forward scanning spectrum to study the nonequilibrium phenomenon. 

In the case of a low Rabi frequency of the probe laser, identical EIT spectra are produced in both scanning directions, as shown in Fig.\ref{Fig.1}(b). As the Rabi frequency of the probe beam increases, more atoms are excited to Rydberg states, which results in an increase in the density of Rydberg atoms. When the probe beam’s Rabi frequency exceeds $\Omega_p$ = 3.5×2$\pi$ MHz, the system crosses a phase-transition threshold of Rydberg atom density, and the bistability immediately emerges. This indicates that the optical intrinsic bistability under Rydberg atoms is caused by interactions between multiple Rydberg atoms and requires that the density threshold is breached. 
The system exhibits two thermodynamic phases: a phase with weak mean-field interactions at low Rydberg-state atomic density (NI phase) and a phase with strong mean-field interactions at high Rydberg-state atomic density (I phase). With strong mean-field interactions, line broadening, frequency shifting, and decoherence phenomena occur \upcite{38,39,40}, which ultimately lead to the emergence of the bistability. By recording the range of noncoincidence of the spectral line in two directions, we are able to plot the phase diagram of the system as shown in Fig.\ref{Fig.1}(c). The red regions denote that the spectral lines from the reverse scan are higher than those from the forward scan, whereas the blue regions indicate that the spectral lines from the forward scan are above those of the reverse scan. It is evident that when the density of the Rydberg atoms does not reach the threshold density, both cases reside in the NI phase, and there are no bistable phenomena in this regime. As the probe beam Rabi frequency approaches the critical point, the noncoincidence phenomenon at the red detuning appears first. As the Rabi frequency continues to increase, the noncoincidence region at both red and blue detunings gradually broadens. The noncoincidence region indicates the boundary between the two phases. Inside the boundary, both directions are in the I phase, and the rest of the positions are in the NI phase. The image is divided into two regions: region (1) represents the area where the mean-field interaction between atoms is weak, and the atomic density has not reached the threshold. At this point, EIT spectral lines are produced. Region (2) represents the atomic density region where bistability can be stably produced. In this region, the changes in bistability are closely related to the Rabi frequency of the probe light. 

The spectral shift($\Delta$) diagram in Fig.\ref{Fig.1}(d) shows a change in the square of the probe beam Rabi frequency. Above the threshold, as the Rabi frequency increases, the spectral shift gradually increases, and the magnitude of shift is linearly related to the square of the Rabi frequency of the probe beam. The obtained results closely relate to previous experimental observations \upcite{15}, indicating the existence of noncoincidence regions at frequencies where red and blue detuning occurs. These regions gradually expand with increasing Rabi frequency of the probe light, demonstrating a linear relationship between the frequency shift and the square of the Rabi frequency. However, it remains to be determined whether external perturbations, such as microwave, electric, or optical fields, will induce nonequilibrium dynamics consistent with earlier findings or give rise to new physical phenomena.

\subsection*{Bistability in AT splitting}
\noindent Following our observation of the bistable phenomenon in the EIT spectrum, we proceed to investigate the bistable situation in the EIT-AT (Autler-Townes) spectrum (details in Methods section). When we observe the spectrum at the resonance point of the microwave frequency, the AT splitting spectrum still exhibits a frequency shift and bistable phenomena. The observed bistable behavior in the EIT-AT spectrum is attributed to the strong mean-field interaction between Rydberg atoms, but it presents differences from the bistability observed in the EIT spectrum. Initially, our intention was to conduct experiments at microwave frequencies resulting in splitting peaks with same height. However, the AT splitting of equal heights obscures the phenomenon of the bistability. To clearly identify the nonequilibrium phase transition region, we varied the frequency of the microwave field and observed the effects under conditions of red and blue detuning relative to the frequency spacing between two Rydberg states. To facilitate this measurement, we designate the microwave that yielded equal-height AT splitting as the zero point when the EIT spectrum had not exhibited bistable behavior. 

\begin{figure}
    \centering
    \includegraphics[width=1 \linewidth]{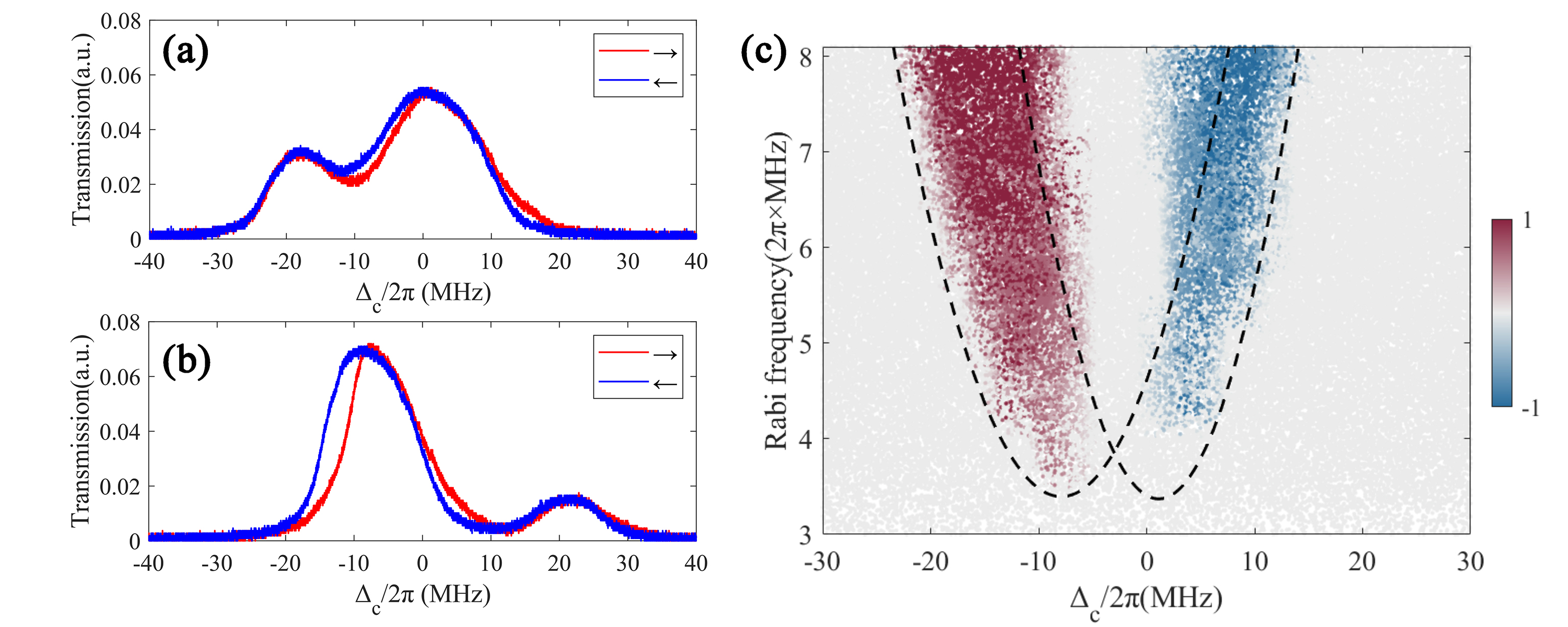}
    \caption{(a) Transmission spectrum with the microwave frequency blue detuning of 90 MHz. (b) Transmission spectrum with the microwave frequency red detuning of 240 MHz. (c) Color map showing the transmission difference of AT splitting when the microwave frequency is red-detuned by 300 MHz.}
    \label{Fig.2}
\end{figure}

The Fig.\ref{Fig.2}(a) demonstrates that when the microwave field is adjusted to a blue detuning of 90 MHz relative to the frequency spacing between two Rydberg states, there is a discrepancy between the reverse scan spectrum and the forward scan spectrum. Moreover, at the center of the splitting peaks, the reverse scan spectrum exhibits higher intensity compared to the forward scan spectrum. The noncoincidence phenomena are stronger at the red detuning of the spectral lines. When the microwave field was set to have a 240 MHz red detuning relative to the frequency spacing between two Rydberg states, the noncoincidence phenomena at the red and blue detuning of coupling beam are mainly concentrated at the higher splitting peak on the red-detuned side (Fig.\ref{Fig.2}(b)). The nonequilibrium phase transition region on the left hand side is redshifted overall compared with the microwave’s blue detuning region. In our experiment, the application of low-intensity microwaves combined with specific detuning resulted in a non-equilibrium phase transition region solely at the position of the higher peak. Therefore, we focused on analyzing the highest peak in subsequent studies as it represents distinctive characteristics of this nonequilibrium phase transition region. As microwave intensity increases and frequency varies, more intricate effects emerge between split peaks \upcite{41}.

 Based on the previously defined microwave zero frequency point, we plot the system phase diagrams at microwave frequency red detuning of 300 MHz (Fig.\ref{Fig.2}(c)). We set the Rabi frequency of probe beam to 6.48×2$\pi$ MHz. As with the previous observations, the density of the Rydberg atoms increased gradually as the Rabi frequency of the probe beam increased, leading to bistable phenomena when the Rydberg atomic density threshold was reached. However, unlike in EIT, the entire nonequilibrium phase transition region is redshifted. The noncoincidence regions of the two detuned AT splittings were significantly smaller than those in EIT, and the range of noncoincidence regions expanded with the increase of the probe beam Rabi frequency. Similarly, in the spectral line plot for the AT splitting, the spectral line of the forward scan still exhibited a peak higher than that of the reverse scan, but this phenomenon only occurred for the spectra with the microwave frequency that was being red-detuned (Fig.\ref{Fig.2}(b)). 

\subsection*{Influence of different detunings}
\begin{figure}
    \centering
    \includegraphics[width=1\linewidth]{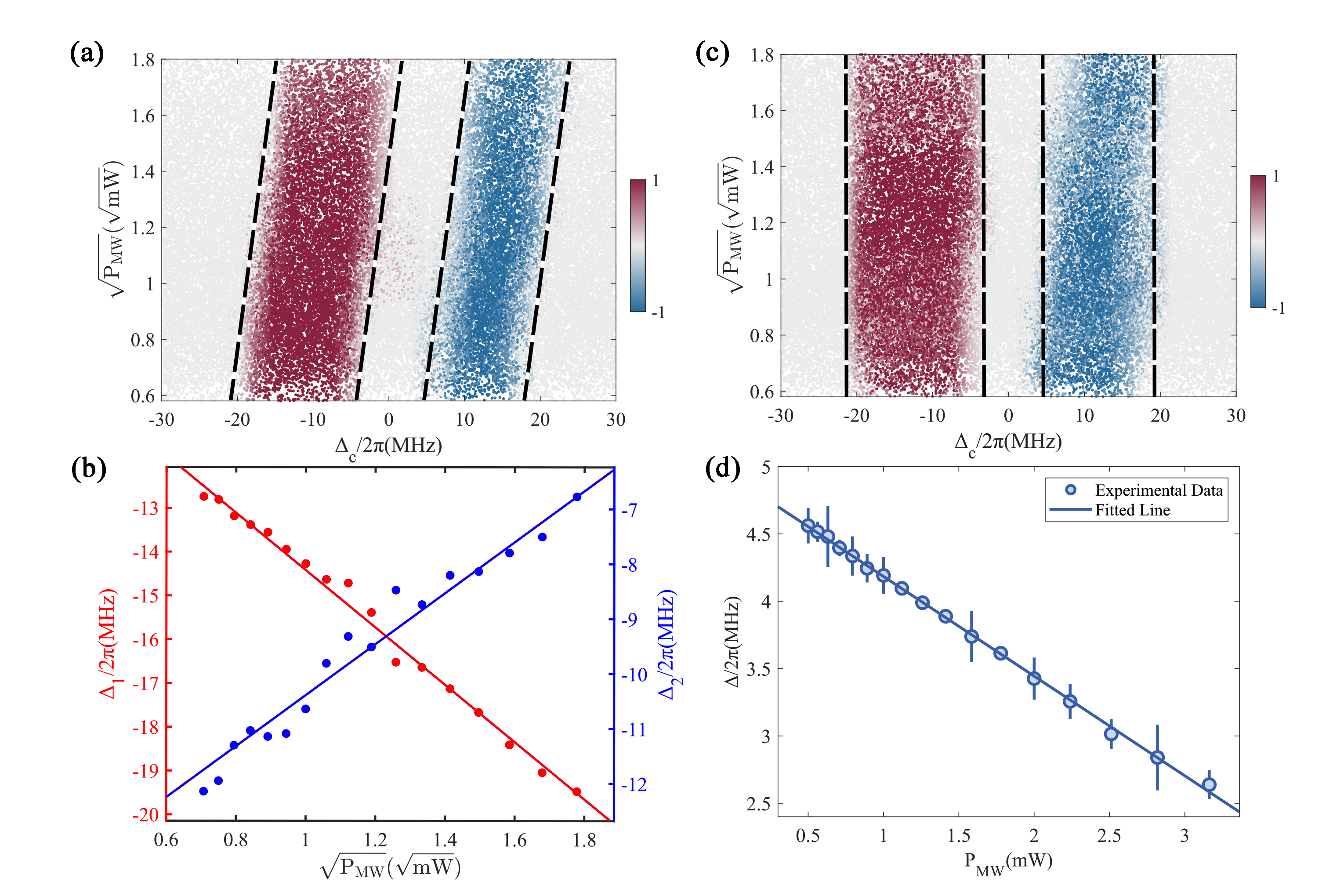}
    \caption{ (a) Color map of microwave intensity versus coupling light detuning at a microwave frequency blue detuning of 300 MHz. Color map of the nonequilibrium phase transition region as a function of microwave power reveals a linear shift in the nonequilibrium phase transition region with the square root of power. (b) The relationship plot between the frequency shift of the center point of bistability at the coupled light red detuning position and the square root of microwave power. The red (blue) segment (red for $\Delta_1$, blue for $\Delta_2$) in the figure represents the relationship at a microwave frequency red (blue) detuned for 300 MHz. The blue segment corresponds to the (a). (c) Color maps showing the transmission difference of EIT when the microwave frequency is red-detuned by 1,500 MHz. (d) Frequency shift of the EIT spectral line versus the microwave power when the microwave frequency is far-detuned and the AT splitting cannot occur. The error bars represent the standard deviation of the data. }
    \label{Fig.3}
\end{figure}

\noindent Subsequently, we investigate the changes in the nonequilibrium phase transition region when varying the microwave power at microwave frequency blue detuning of 300 MHz, as shown in Fig.\ref{Fig.3}(a). It can be observed that the nonequilibrium phase transition region correlates with the square root of microwave power in a linear fashion. To ascertain the accuracy of microwave power determination using this method, we conduct measurements with microwave frequency red and blue detunings of 300 MHz. Center points of the nonequilibrium phase transition region ({\it i.e.}, the geometrical centers of the bands outlined in Fig.\ref{Fig.3}(a)) at the coupled light red detuning position {\it vs} the square root of the microwave power are depicted in Fig.\ref{Fig.3}(b). Notably, the center values demonstrate a linear relationship with the square root of microwave power.

This phenomenon suggests the possibility of a microwave measurement technique that does not require precise microwave wavelengths. Previously, microwave measurement using EIT-AT spectra involved generating two equally high split peaks in the EIT-AT spectrum at the microwave resonance position. The microwave intensity was then determined by measuring the frequency difference between these two peaks. Now, we can focus solely on the nonequilibrium phase transition region that appears in the higher of the split peaks. By observing the changes in this region, we can measure the microwave intensity more efficiently. The spectra of ordinary AT-split exhibit a Gaussian profile similar to that of EIT spectra. However, in the case of bistability, the AT-split spectra show a non-overlapping region during both forward and reverse scans. In this specific region, the spectral line displays a steeper slope compared to the typical Gaussian profile, as indicated by the red line on the left side of Fig.\ref{Fig.2}(b). In this region, the microwave-induced shift in the spectral line is more pronounced than that in the conventional AT splitting. And it underscores the influence of microwaves coupled to two Rydberg states on the bistable behavior under the AT splitting. Eventually, we can accurately measure the power of the non-resonant microwave.

When the microwaves resonate with the two Rydberg states of cesium atoms, an EIT-AT effect occurs. When the microwave frequency is far-detuned, it does not produce equally high AT splitting peaks. However, due to the AC Stark effect, the EIT spectrum experiences a frequency shift. When microwaves are resonant, the presence of equally high AT split peaks poses challenges in observing the bistable region and affects its intrinsic properties. Therefore, we conduct experiments using detuned microwaves to mitigate these issues. However, microwave detuning introduces difficulties in distinguishing whether changes in the bistable region are caused by frequency shifts due to the AC Stark effect or if microwave interference alters mean-field interactions between atoms. To address this concern, we further adjust the microwave detuning and observe the system under far-detuning conditions. If similar changes occur in the bistable region under far-detuning conditions, it indicates that these changes are solely attributed to frequency shifts caused by the AC Stark effect and are unrelated to alterations in the second Rydberg state level. Conversely, if different changes occur, they can be attributed to microwaves coupling with the second Rydberg state. Using a microwave with a red detuning of 1500 MHz relative to the frequency spacing between the two Rydberg states for the experiment, we observed that at this microwave frequency, the spectrum exhibited only a single transparency window instead of two split peaks.

As shown in Fig.\ref{Fig.3}(c), we confirm that a far-detuned microwave field does not cause a frequency shift for the nonequilibrium phase transition region. However, the AC Stark effect still affects the overall frequency shift of the spectral lines (Fig.\ref{Fig.3}(d)). The frequency shift of the EIT spectral line exhibits a linear relationship with the microwave power. The fact that far-detuned microwaves do not affect the position of the nonequilibrium phase transition region indicates that the frequency shift of the center point in the nonequilibrium phase transition region observed in the experiment originates solely from the resonant interaction with the microwave field. This further enhances the feasibility of measuring microwave power \upcite{42,43,44,45} using the nonequilibrium phase transition region. The EIT-AT effect is difficult to observe at low microwave powers. However, the size of the nonequilibrium phase transition region can be effectively increased by adjusting the Rabi frequency of the probe beam, which makes the resulting frequency shift easier to observe relative to the frequency shift that results from regular AT splitting.

\subsection*{Rate of change of the transmission spectrum}
\begin{figure}
    \Scentering
    \includegraphics[width=1\linewidth]{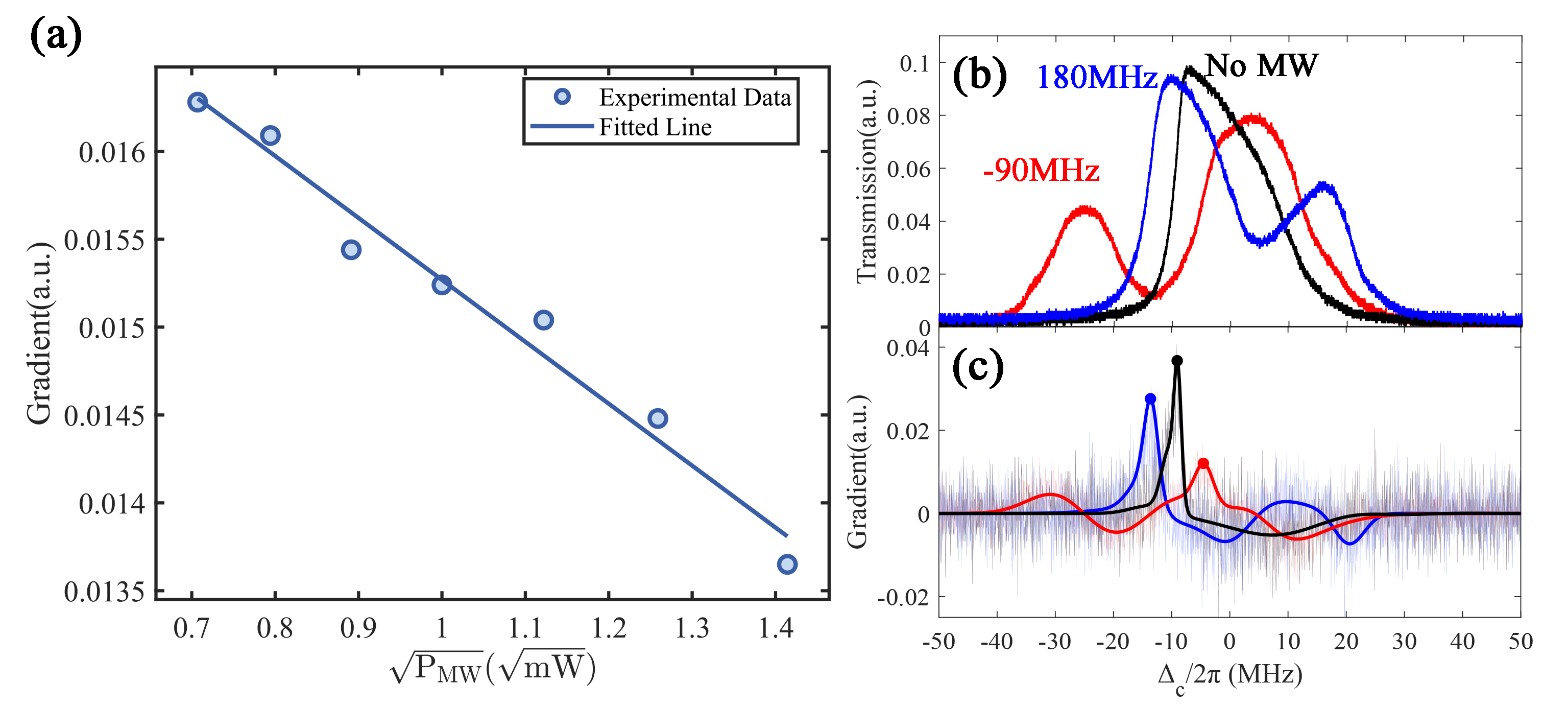}
    \caption{(a) The maximum rate of change (gradient) in the forward transmission spectrum as a function of the square root of the microwave power, with a microwave frequency detuning of 300 MHz.(b) The red, blue, and black spectra represent the forward scan transmission spectra with microwave frequency detunings of -90 MHz, 180 MHz, and no microwave, respectively. (c) The corresponding rate of change (gradient) of the transmission spectra shown in (b). The red, blue, and black points represent the highest points of the three lines, which are the maximum values of the spectral gradient in Figure b. These points are red (-0.455, 0.012), blue (-13.64, 0.0276), and black (-9.05, 0.0367).}
    \label{Fig.4}
\end{figure}

\noindent Due to changes in microwave power and frequency, the changes of nonequilibrium phase transition region not only involve a shift in position but also alter the shape. By modifying the shape of the critical region, we can simulate the dynamic process of non-equilibrium phenomena with two stable states. The focus of nonequilibrium phenomena research is on the transition process between steady states. In the Rydberg atom system, the critical region represents the transition between the I phase and the NI phase. We focus on the rate of change (gradient) of the transmission spectrum with respect to the coupling light Rabi frequency. As illustrated in Figure \ref{Fig.4}(a), we investigated the influence of microwave power on the maximum rate of change of the spectral lines (Gradient) for the microwave frequency blue detuning. When adjusting microwave power and observing spectral line changes, we noted that the maximum rate of change varied with changes in microwave power. Notably, as the square root of microwave power increased gradually, the rate of change of the spectral lines gradually decreased. The change in gradient within the designated microwave power range for our experiment exhibits an approximately linear relationship.
Besides, as depicted in Fig.\ref{Fig.4}(b),(c), we manipulate the microwave frequency and observe the spectral line changes during forward scans. In the absence of microwave (black line in the figure (b)), the spectral lines display an asymmetric shape. Through data analysis, we determine the rates of change at each point, with a maximum positive rate of change reaching 0.036 (The peak of the black line in the figure (c)). The position where the spectral line rises corresponds to the non-overlapping region of the spectra in the two scanning directions. Upon microwave application, both microwave frequency red and blue detunings lead to distinct alterations in the maximum rate of change of the spectral lines, the nonequilibrium phase transition region of the spectra gradually becomes smoother. When the microwave frequency is red-detuned by 90 MHz, the maximum gradient of the spectral line is 0.012. When the microwave frequency is blue-detuned by 180 MHz, the maximum gradient of the spectral line is 0.0276.

The microwave-coupled Rydberg atomic system provides significant assistance in simulating nonequilibrium phenomena. Microwave intervention can simulate various external disturbances under nonequilibrium conditions. For example, the bistabie phenomenon in Rydberg atom EIT can be used to simulate the epidemic model, where the proportion of infected individuals and population size exhibit non-equilibrium behavior (refer to the supplementary information for specific details). The microwave field can simulate the effects of external disturbances on the model, such as changes in disease transmissibility after vaccination. Similarly, by altering microwave intensity, frequency, and other parameters, various disturbances in nonequilibrium models can be simulated. These simulations can extend to perturbations in other models, enhancing our understanding and modeling capability of external disturbances' impacts during the evolution of nonequilibrium systems. With further confinement of the experimental conditions, they can simulate more observable nonequilibrium phenomena in daily life, which holds profound significance for understanding the complex systems.

\section*{DISCUSSION}
\noindent In this work, we have studied a nonequilibrium phase transition driven by microwave and light, as well as the intrinsic optical bistability in a vapor of cesium Rydberg atoms. We have obtained the phase diagram in the vicinity of the system’s critical point, where a phase transition is observed. Both EIT and AT splitting spectra are used to study nonequilibrium phase transitions via the inducement of bistable states. The observation of nonequilibrium phenomena in the bistable spectral lines, influenced by microwaves fields, suggests the potential use of thermal Rydberg atoms to simulate disturbances in nonequilibrium phenomena encountered in daily life. This can further advance the understanding and manipulation of quantum simulations of nonequilibrium phenomena. Additionally, it offers significant assistance in addressing issues related to resolving nonequilibrium problems.

The far-detuned microwave does not cause changes in the bistable spectra. This indicates that the bistable spectra exhibit robustness against non-resonant microwaves. Under slight microwave frequency detuning, AT splitting exhibits significant phase transitions and shows the microwave properties. The characteristics of bistable spectra under AT splitting provides a novel approach to microwave measurement. In addition, nonequilibrium dynamics are sensitively dependent on near-resonant microwave fields, providing a tool to probe microwave fields through the dynamics. 
\justifying

\section*{METHODS}
\subsection*{Experimental setup}
\begin{figure}
    \centering
    \includegraphics[width=1\linewidth]{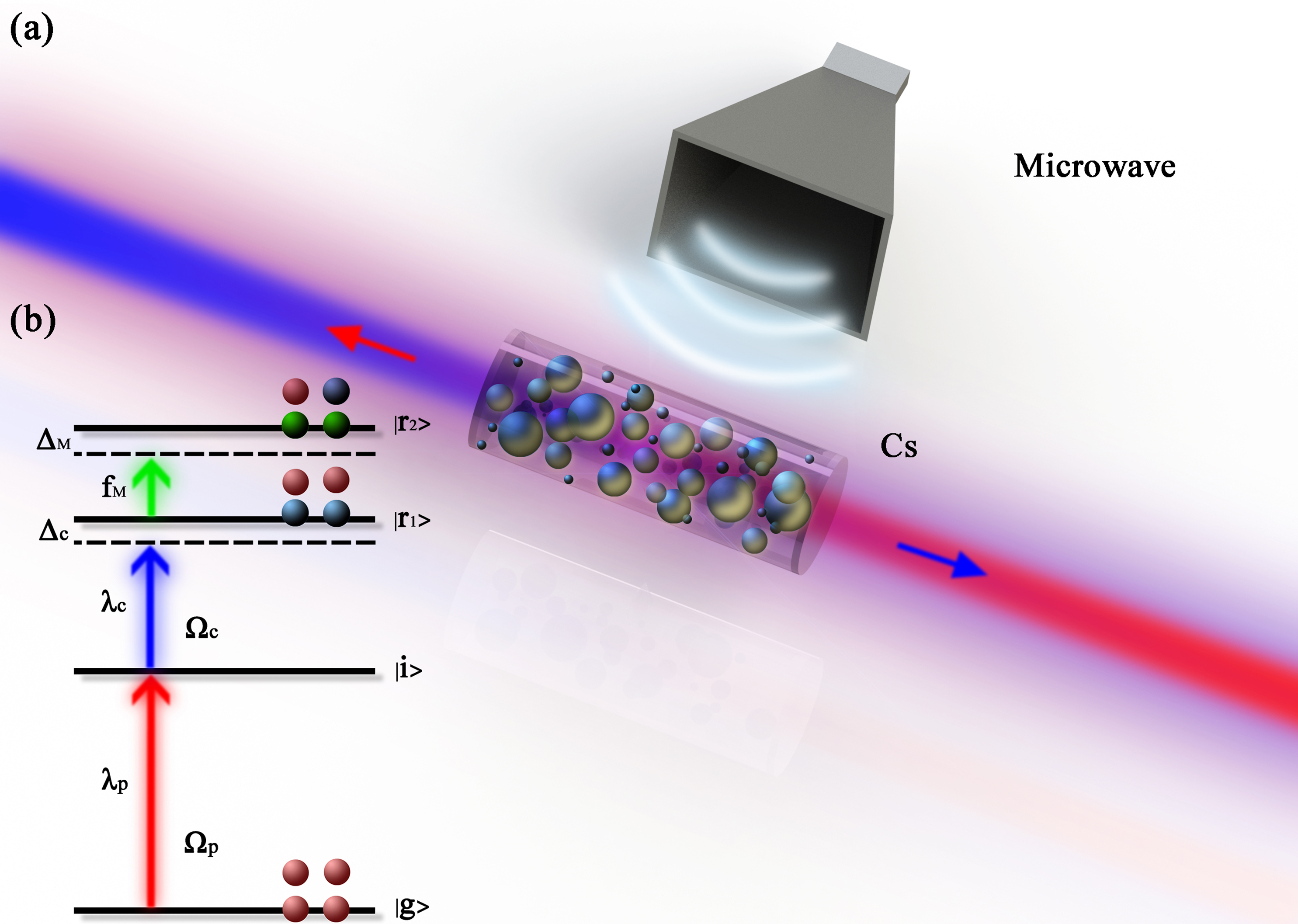}
    \caption{(a) Schematic of the experimental setup. A probe beam is overlapped with a counter-propagating coupling beam. The transmitted signal of the probe beam is collected by a photodetector. The microwave horn is aligned with the cesium atomic vapor cell. (b) Excitation scheme to Rydberg states in cesium. }
    \label{Fig.5}
\end{figure}

\noindent The atoms are initially excited by laser through a two-photon excitation process, inducing strong interactions and driving them to the desired Rydberg states. This results in the formation of an ensemble of Rydberg atoms, as shown in Fig.\ref{Fig.5}(b) (similar to \cite{46}). Given the two-level structure of atoms excited to a Rydberg state, the full width at half maximum (FWHM) of the absorption line is a combination of the natural linewidths of the upper and lower levels. However, when we select the three-level structure, the population of the ground state is coherently pumped from the intermediate state by a coupling field, and the FWHM is only determined by the natural linewidth of the final excited state \upcite{47}. Therefore, this nonequilibrium phenomenon in EIT exceeds the influence of spectral linewidths. The spectral linewidth is narrow enough to clearly observe the nonequilibrium phase transition.

A probe beam propagates through the cesium vapor cell, exciting cesium atoms from the ground state to the intermediate state, and then the transmitted beam from the cesium vapor cell enters the photodetector, as shown in Fig.\ref{Fig.5}(a). We used an 852 nm semiconductor laser (DL) to provide the probe beam, the level transition is from $\mid$g$\rangle$ = $\mid$6$S_{1/2}$, $F$ = 4$\rangle$ to $\mid$i$\rangle$ = $\mid$6$P_{3/2}$, $F$ = 5$\rangle$. The beam is locked by the polarized absorption spectrum at the resonance frequency of the transition, and the Rabi frequency is $\Omega_p$. The coupling beam, which has the wavelength $\lambda_c$ = 510 nm and the Rabi frequency $\Omega_c$, generated by a second harmonic generator (SHG), propagates in directions opposite to that of the probe beam, and it is resonant with the intermediate state and the first Rydberg states. The level transition by this beam is from $\mid$i$\rangle$ = $\mid$6$P_{3/2}$, $F$ = 5$\rangle$ to $\mid$$r_1$$\rangle$ = $\mid$$nD_{5/2}$$\rangle$. The microwave field couples the levels $\mid$$r_1$$\rangle$ = $\mid$$nD_{5/2}$$\rangle$ and $\mid$$r_2$$\rangle$ = $\mid$$(n+1)P_{3/2}$$\rangle$. 

\subsection*{AT splitting}
\noindent We use an analog signal microwave generator to propagate microwave of a specific frequency into the cesium vapor cell; the microwave horn is maintained at the same height as the cesium vapor cell, and the polarization direction of the microwave field is parallel to the direction of the optical polarization entering the vapor cell. Coupling between Rydberg atoms and the microwave field of various frequencies can induce coherence effects, {\it e.g.}, the Autler–Townes splitting caused by near-resonant RF fields, and the AC Stark shifts caused by far-detuned RF fields. When the microwave field frequency is resonant with the energy spacing between the Rydberg levels, the transmission peak (representing the transparent window in the absorption line profile) splits into two symmetric peaks. By measuring the splitting width of these two peaks, we can determine the intensity of the microwave field. By applying modulation techniques, various information such as the frequency, phase, and waveform of the microwave field can be obtained \upcite{48,49,50,51}. 

\medskip
\noindent\textbf{Data Availability}\par
\justifying\noindent Data underlying the results of this study are available from the authors upon request.

\medskip
\noindent\textbf{Code Availability}\par
\justifying\noindent The codes used to generate data for this paper are available from the corresponding author upon reasonable request.

\medskip
\noindent\textbf{Acknowledgements}\par  
\noindent
\justifying This work is supported by the Innovation Program for Quantum Science and Technology (Grant No. 2021ZD0302103), the National Natural Science Foundation of China (Grant Nos. 62325505, 62020106014, 62175140, 62475138, 92165106, 12104276); the Applied Basic Research Project of Shanxi Province (Grant No. 202203021224001); the Shanxi Province Graduate Student Research Innovation Project (Grant No. 2024KY105).

\medskip
\noindent\textbf{Author Contributions} \par
\noindent
\justifying Z.H. Zhang and J. Z. Wu conceived the original idea. Z. H. Zhang, S. X. Han and Z. Y. Zhang  performed the measurements and analysis of the results. Z. H. Zhang and Z. Y. Zhang designed and simulated the device. Z. H. Zhang, Z. Y. Zhang and Y. Q. Zhang developed the idea and carried out theoretical simulations of the system. Z. H. Zhang, S. X. Han and G. Q. Zhang processed the experimental data. Zhang, Z. H. and J. Z. Wu wrote the manuscript. J. Z. Wu, V. Sovkov, L.J. Zhang and W. B. Li edited the manuscript. W. L. Liu, Y. Q. Li, L. T. Xiao, S. T. Jia, J. Ma provided guidance during the project. J. Z. Wu and J. Ma supervised the whole project. All authors contributed to the discussions and interpretations of the results.

\medskip
\justifying\noindent\textbf{Competing Interests}\par
\noindent The authors declare no competing of interests.

\medskip

%


\end{document}